\documentclass[preprint,amsmath,amssymb,aps]{revtex4-1}
\pdfoutput=1
\usepackage[pdftex]{graphicx}
\usepackage{dcolumn}
\usepackage{bm}
\usepackage{color}

\begin{document}
\renewcommand{\theenumi}{\alph{enumi}}


\title{
AdS/QCD approach to the scale-invariant extension of the standard model with
a strongly interacting hidden sector
} 

\author{Hisaki Hatanaka}
 \email{hatanaka@kias.re.kr}
\affiliation{Quantum Universe Center, KIAS, Seoul 02455, Korea}
\author{Dong-Won Jung}%
 \email{dongwonj@korea.ac.kr}
\affiliation{Department of Physics, Korea University, Seoul 136-713, Korea}
\author{Pyungwon Ko}%
 \email{pko@kias.re.kr} 
\affiliation{School of Physics and Quantum Universe Center, KIAS, Seoul 02455, Korea} 





\begin{abstract}
In this paper, we revisit a scale-invariant extension of the standard model (SM) 
with a strongly interacting hidden sector within AdS/QCD approach. 
Using the AdS/QCD, we reduce the number of input parameters to three, {\it i.e.}  
hidden pion decay constant, hidden pion mass and $\tan\beta$ that is defined as the 
ratio of the vacuum expectation values (VEV) of the singlet scalar field and the 
SM Higgs boson. As a result, our model has sharp predictability. We perform 
the phenomenological analysis of the hidden pions which is one of the dark matter (DM)
 candidates in this model. With various theoretical and experimental constraints 
we search for the allowed parameter space and find that both resonance and 
non-resonance solutions are possible. Some typical correlations among various 
observables such as thermal relic density of hidden pions, Higgs boson signal strengths 
and DM-nucleon cross section are investigated. We provide some benchmark points 
for experimental tests.
\end{abstract}

\pacs{Valid PACS appear here}
\maketitle


\section{Introduction}\label{sect-intro}

Although the SM-like Higgs boson has been discovered at the Large Hadron Collider (LHC) 
\cite{Aad:2012tfa,Chatrchyan:2012xdj},  
there are still a number of questions that call for physics beyond the SM (BSM): 
(i) the origin of the mass of Higgs particle or the origin of weak scale, 
(ii)  the nature of non-baryonic dark matter (DM),  (iii) the origin of neutrino masses and
mixing, (iv) matter-antimatter asymmetry of the universe, to name a few.

The first question is often phrased as hierarchy problem, that addresses why
the electroweak (EW) scale $v_H=246$ GeV is much smaller than the Planck scale. 
One of the nice ways to understand this is through quantum dimensional transmutation,
which explains why the proton mass is much suppressed compared with 
the Planck mass in Quantum Chromodynamics (QCD) \cite{Hill:2002ap,Wilczek:2005ez}. 
Technicolor (TC) provides an answer in this way, but the naive version of it is strongly 
disfavored by the electroweak precision test (EWPT) \cite{Peskin:1990zt}.

Since the observation of W. Bardeen \cite{Bardeen:1995kv},  softly broken scale invariance 
has been considered as a possible solution for the hierarchy problem. 
If the model is scale invariant at classical level, no dimensionful  parameters are allowed 
and the scale symmetry is broken only logarithmically  through scale-anomaly. 
Many authors have studies this type of models where the EW symmetry is dynamically 
broken via dimensional transmutation in the hidden sector with new confining strong interactions 
\cite{Hur:2007uz,
Ko:2008ug,
Ko:2009zz,
Ko:2010rj,
Hur:2011sv,
Heikinheimo:2013fta,
Heikinheimo:2013xua,
Holthausen:2013ota,
Jung:2014zga,
Salvio:2014soa,
Kubo:2014ova,
Kubo:2014ida,
Schwaller:2015tja,
Ametani:2015jla,
Kubo:2015joa,
Haba:2015qbz}, 
or Coleman-Weinberg mechanism 
\cite{Meissner:2006zh,
Foot:2007as,
Meissner:2007xv,
Hambye:2007vf,
Foot:2007iy,
Iso:2009ss,
Iso:2009nw,
Holthausen:2009uc,
AlexanderNunneley:2010nw,
Ishiwata:2011aa,
Lee:2012jn,
Okada:2012sg,
Iso:2012jn,
Gherghetta:2012gb,
Das:2011wm,
Carone:2013wla,
Khoze:2013oga,
Farzinnia:2013pga,
Antipin:2013exa,
Hashimoto:2014ela,
Hill:2014mqa,
Guo:2014bha,
Radovcic:2014rea,
Binjonaid:2014oga,
Allison:2014zya,
Farzinnia:2014xia,
Pelaggi:2014wba,
Farzinnia:2014yqa,
Foot:2014ifa,
Benic:2014aga,
Guo:2015lxa,
Oda:2015gna,
Fuyuto:2015vna,
Endo:2015nba,
Plascencia:2015xwa,
Hashino:2015nxa,
Karam:2015jta,
Ahriche:2015loa,
Wang:2015cda,
Haba:2015nwl,
Ghorbani:2015xvz,
Helmboldt:2016mpi,
Jinno:2016knw,
Ahriche:2016cio,
Ahriche:2016ixu,
Das:2016zue,
Khoze:2016zfi}.

Some of the present authors have proposed a scale-invariant extensions of the SM with 
a strongly interacting hidden sector, namely hidden QCD models 
\cite{Hur:2007uz,Ko:2008ug,Ko:2009zz,Ko:2010rj,Hur:2011sv,Hur:2011sv}. 
At the classical level,  the scale invariance is imposed so that all dimensionful parameters 
are forbidden in the classical Lagrangian. The Higgs mass term arises at quantum level 
through the dimensional transmutation driven by asymptotically free gauge theories
 in the hidden sector. Hidden sector couples to the Higgs through singlet scalar field only 
and there are stable or long-lived particles (lightest hidden mesons and hidden baryons) 
that can make good DM candidates.  In those works, hidden QCD sector was studied 
in the chiral effective Lagrangian approach and non-perturbative parameters were estimated 
by naive dimensional analysis.  Then the same model was analyzed in  the 
Nambu--Jona-Lasinio (NJL) approach in Ref.~\cite{Holthausen:2013ota,Kubo:2014ida}. 

In this paper, we consider the same model using another approximation method, the 
AdS/QCD \cite{DaRold:2005zs,DaRold:2005vr}, in order to analyze 
non-perturbative strong dynamics in the hidden QCD models.
First we reformulate the hidden QCD sector in terms of the linear sigma model,
in which the sigma and pi mesons are effective degrees of freedom. 
We consider a linear sigma model coupled with a scale-invariant Higgs-singlet 
sector and analyzed the masses and mixing between the SM Higgs boson, a singlet 
scalar messenger and the sigma meson. 
In the AdS/QCD we successfully reduce the number of free parameters by matching 
the mass spectra of the lightest scalar, vector and axial vector mesons.

Next, we apply this model to dark matter phenomenology.
In our model, since the hidden quarks do not couple to any U(1) gauge fields,
the hidden pions cannot decay through the U(1) anomaly and are found to be stable lightest particles coupling weakly with the SM fields. Hence the hidden pions become candidates of the weakly interacting massive particle (WIMP) DM.
With the free parameters reduced by the AdS/QCD, we identify the parameter space 
that satisfies the recent observations.  Then we study the distinctive features of 
the allowed parameter region and also some typical correlations among various 
observables. We address on the possible signatures of the model that can be further 
scrutinized in the future experiments such as LHC Run-II, ILC  and so forth. 

This paper is organized as follows.
In Sec.~\ref{sect-model}, we revisit the original hidden QCD models 
\cite{Hur:2007uz,Hur:2011sv} by reformulating the hidden QCD sector with the linear 
sigma model. Then in Sec.~\ref{sec:qcdqcd}, we apply the idea of 
the AdS/QCD to the linear sigma model described in Sec.~\ref{sect-model}. 
In Sec.~\ref{sect-dm}, numerical results on the Higgs and the dark matter phenomenologies 
 are presented.  Then Sec.~\ref{sect-summary} is devoted to summary and discussions.

\section{The Model}\label{sect-model}
Scale-invariant extension of the SM with a strongly interacting hidden sector 
contains the SM fields  plus a singlet scalar $S$ and a scale-invariant hidden QCD sector 
\cite{Ko:2008ug,Ko:2009zz,Ko:2010rj,Hur:2011sv}. 
The corresponding Lagrangian is given by 
\begin{eqnarray}
{\cal L} 
&=&
{\cal L}_{\rm SM} (\mu_H^2 = 0) 
+ \frac{1}{2} (\partial_\mu S)^2 
 -  \frac{\lambda_S}{8} S^4
 + \frac{\lambda_{HS}}{2} H^\dag H S^2 - \frac{1}{2} Y^{ij}_{N_R} S \overline{N_{Ri}^c} N_{Rj} 
 \nonumber \\
&&
- 
 \left(  Y_N^{ij} \overline{N_{Ri}} \widetilde{H} l_{Lj} + H.c.  
 \right)
 - \frac{1}{2} \text{tr}\, {G}_{\mu\nu} G^{\mu\nu} 
+ \sum_{k=1}^{N_{h,f}} \overline{Q_k} (i\gamma^\mu D_\mu - \lambda_{Q,k} S) Q_k,  
\end{eqnarray}
where $\mu_H$ is the mass parameter of the SM Higgs boson. We have replaced all the mass
parameters (the Higgs boson mass, the RH neutrino masses and the current quark masses of 
hidden-sector quarks) by real singlet scalar operators $S$ or $S^2$ following the idea of classical
scale invariance.  $G_{\mu\nu}$  is the field strength of the hidden QCD with $SU(N_{h,c})$ 
gauge symmetry.  The SM singlet scalar $S$ couples to the hidden-sector quarks $Q_f$ 
through the Yukawa interaction.  Since there are no dimensionful parameters in the 
Lagrangian, this system is scale-invariant at the classical level.
At quantum level and low-energy scale, the hidden-QCD quarks can condensate.
Such condensates $\langle \bar{Q}Q \rangle$ induce a linear term in $S$.
Then the potential of $S$ can be tilted and $S$ can develop a VEV.
 The VEV of the singlet scalar generates a Higgs boson mass term $-\frac{\lambda_{HS}}{2} 
 \langle S\rangle^2 H^\dag H$, as well as the RH neutrino masses and the current quark 
masses of the hidden-sector quarks.  Thus all the mass scales in this model are generated 
by $\langle S \rangle$, which is a result of non-perturbative dynamics in the strongly interacting 
hidden sector.   For this to happen, we assume that $\lambda_{HS} > 0$ so that non-zero 
$\langle S \rangle$ triggers the electroweak symmetry breaking.

Hereafter we consider the case in which $N_{h,c}=3$ and $N_{h,f}=2$, for which we can 
use the known results from the hadronic system with $\pi, \rho$ and $\sigma$ mesons.
Then $\lambda_Q = \text{diag}(\lambda_{Qu},\lambda_{Qd})$ and for simplicity we assume the hidden quarks have isospin symmetry $\lambda_{Qu} \sim \lambda_{Qd}$.
In such a case the low-energy effective theory of the hidden QCD is described by 
the pi meson triplets and the sigma meson. 
It would be written in the form of a linear sigma model
\begin{eqnarray}
{\cal L}_{L\sigma M}
&=& \frac{1}{2} (\partial_\mu \Sigma)^2  + \frac{1}{2} (\partial_\mu \pi)^2
 + \frac{\lambda_{\sigma}}{4} (\Sigma^2  + \pi^2)^2
- \frac{\mu_\sigma^2}{2} (\Sigma^2  + \pi^2)
 - m_{S\sigma}^2 S\Sigma,
\end{eqnarray}
where $\Sigma$ and $\pi$ represents sigma and pi meson fields.

We parameterize the VEVs and fluctuations of scalars as
\begin{eqnarray}
H = \frac{1}{\sqrt{2}} \begin{pmatrix}0  \\ v_H + h \end{pmatrix},
\quad
S = v_S + s,
\quad
\Sigma = v_\sigma + \sigma. 
\end{eqnarray}
To minimize the potential energy
\begin{eqnarray}
V(v_H, v_S,v_\sigma)
 &=& \frac{\lambda_{H}}{8} v_H^4 - \frac{\lambda_{HS}}{4} v_H^2 v_S^2
+ \frac{\lambda_S}{8} v_S^4
+ \frac{\lambda_\sigma}{4} v_\sigma^4 - \frac{\mu_\sigma^2}{2} v_\sigma^2
- m_{S\sigma}^2 v_S v_\sigma,
\end{eqnarray}
three minimization conditions $\partial V/\partial v_\phi=0$ ($\phi = H, S, \sigma$) should be satisfied.
These conditions reduce the number of free parameters. Furthermore, two parameters $\lambda_\sigma$, $\mu_\sigma$ are traded with the pion mass $M_\pi$ and a sigma meson mass parameter $M_{\sigma\sigma}$.
Hence the scalar mass matrix ${\cal L} \supset -\frac{1}{2} (h,s,\sigma){\cal M}(h,s,\sigma)^T$ 
takes the form of
\begin{eqnarray}
{\cal M} &=& 
\begin{pmatrix} 
M_{hh}^2 & M_{hs}^2 & 0 \\
M_{hs}^2 & M_{ss}^2 & -m_{S\sigma}^2 \\
0 & -m_{S\sigma}^2 & M_{\sigma\sigma}^2
\end{pmatrix}
\label{mmatrix}
\end{eqnarray}
with
\begin{eqnarray}
M_{hh}^2 &=& \lambda_H v_H^2 = \lambda_{HS}v_H^2 \tan^2\beta,
\\
M_{hs}^2 &=& -\lambda_{HS} v_H^2 \tan\beta,
\\
M_{ss}^2 &=& \lambda_{HS} v_H^2 \left( 1 + \frac{3 M_\pi^2 F_\pi^2}{\lambda_{HS} \tan^2\beta v_H^4}\right),
\\
-m_{S\sigma}^2 &=& -\frac{M_\pi^2 F_\pi}{v_S},
\end{eqnarray}
where $\tan\beta \equiv v_S/v_H$ and $F_\pi \equiv v_\sigma$.
Since a off-diagonal part $M_{hs}^2$ satisfies
$M_{hs}^2 = - M_{hh}^2/\tan\beta$, the Higgs-singlet mixing can be large when $\tan\beta$ is small.
$\lambda_\sigma$ and $\mu_\sigma$ are traded with pion mass $M_\pi$ and sigma meson mass $M_{\sigma\sigma}$ by
\begin{eqnarray}
\lambda_\sigma = \frac{M_{\sigma\sigma}^2 - M_\pi^2}{2F_\pi^2},
\quad
\mu_\sigma^2 = \frac{M_{\sigma\sigma}^2 - 3 M_\pi^2}{2}.
\label{positivemu2}
\end{eqnarray}
The couplings $\lambda_H$, $\lambda_S$ are given by
\begin{eqnarray}
\lambda_H = \lambda_{HS} \tan^2\beta,
\quad
\lambda_S = \frac{\lambda_{HS} v_H^4 \tan^2\beta + 2 M_\pi^2 F_\pi^2}{v_H^4 \tan^4\beta},
\end{eqnarray}
where $v_H = 246~\text{GeV}$.
Since one of the physical scalar should be the Higgs boson with mass $M_H=125~\text{GeV}$,
$M_H^2$ is  one of the eigenvalues of ${\cal M}$.
Thus a condition $\det({\cal M} - M_H^2 I_3) = 0$ ($I_3$ is a $3\times3$ unit matrix) yields
\begin{eqnarray}
\lambda_{HS} 
&=& 
\frac{1}{v_S^2} \frac{
3 \xi_\sigma^2 F_\pi^4 M_\pi^2 - F_\pi^2 M_H^2 [3 M_H^2 M_\pi^2 + M_\pi^4 + M_H^2 \xi_\sigma^2 v_H^2 \tan^2\beta] + M_H^6 v_H^2 \tan^2\beta}{
3 \xi_\sigma^2 F_\pi^4 - F_\pi^2 [3 M_\pi^2 M_H^2 + M_\pi^4 + M_H^2 v_H^2 (1 + \tan^2\beta) \xi_\sigma^2] +  M_H^4 v_H^2 (1 + \tan^2\beta)},
\end{eqnarray}
where we have parameterized $M_{\sigma\sigma} = \xi_\sigma F_\pi$.

The mixing matrix is defined as 
\begin{equation}
\left(
\begin{array}{c}
h \\ s \\ \tilde{\sigma}
\end{array}
\right)=
\left(
\begin{array}{ccc}
V_{h0} & V_{h1} & V_{h2} \\
V_{s0} & V_{s1} & V_{s2} \\
V_{\sigma 0} & V_{\sigma 1} & V_{\sigma 2} \\
\end{array}
\right)
\left(
\begin{array}{c}
H \\ H_1 \\ H_2
\end{array}
\right),
\label{mixingmat}
\end{equation}
where $H$ is the SM-like Higgs boson with $M_H=125$ GeV and $H_1, H_2$ are 
extra scalar particles with $M_{H1} < M_{H2}$.

At this stage we have four free parameters: $v_S$, $v_\sigma\equiv F_\pi$, $M_\pi$ and $M_{\sigma\sigma}$ (or $\xi_\sigma$).
To reduce the number of free parameters, in particular, to relate the $M_{\sigma\sigma}$ with $F_\pi$, we use a holographic treatment of the hidden QCD.

\section{AdS/QCD analysis \label{sec:qcdqcd}}
In the AdS/QCD \cite{DaRold:2005zs,DaRold:2005vr}, the hidden QCD sector is 
described by $SU(2)_L \otimes SU(2)_R$ gauge theory on $AdS_5$ space with metric
\begin{eqnarray}
ds^2 &=& a^2(z)(\eta_{\mu\nu}dx^\mu dx^\nu - dz^2) ,
\quad a(z) = \frac{L}{z},
\end{eqnarray}
where $L_0 \le z \le L_1$ and $L$ is the curvature radius of $AdS_5$.
$L_1$ breaks the conformal symmetry in the infrared (IR) regime, while one can take
$L_0$ to be arbitrary small, $L_0 \to 0$. 
The non-perturbative breaking of chiral symmetry is regarded as
the spontaneous breaking of $SU(2)_L \otimes SU(2)_R$ symmetry  by the VEV of the bulk scalar $\Phi$ which is a bi-doublet $(\bm{2}_L, \bm{\bar{2}}_R)$.  
The parity transformation corresponds to the exchange 
$SU(2)_L \leftrightarrow SU(2)_R$.  

Then the 5D bulk Lagrangian is given by
\begin{eqnarray}
S_{5}
&=& \int d^4x \int_{L_0}^{L_1} dy \sqrt{g} M_5 \text{Tr} \left[-\frac{1}{4} L_{MN} L^{MN} -\frac{1}{4} R_{MN} R^{MN} + \frac{1}{2} |D_M \Phi|^2 - \frac{1}{2} M_\Phi^2 |\Phi|^2 \right],
\end{eqnarray}
where $D_M \Phi = \partial_M \Phi + iL_M \Phi - i \Phi R_M$, $M=(\mu,5)$,
and the bulk mass parameter $M_\Phi^2 = -3/L^2$ is chosen so as to relate the bulk scalar 
field   $\Phi$ with the dimension-three operator $\bar{q}q$.   The profile of the VEV 
 is obtained by solving the zero-mode equation of motion.  We have
\begin{eqnarray}
\langle \Phi \rangle \equiv v(z) = c_1 z + c_2 z^3,
\quad v \propto  1_{2\times2},
\end{eqnarray}
where $c_1$ and $c_2$ can be written in terms of the value of $v$ at boundaries
\begin{eqnarray}
c_1 = \frac{\tilde{M}_q L_1^3 - \xi L_0^3}{L L_1 (L_1^2 - L_0^2)},
\quad
c_2 = \frac{\xi - \tilde{M}_q L_1}{L L_1(L_1^2 - L_0^2)},
\end{eqnarray}
from boundary conditions
\begin{eqnarray}
\tilde{M}_q = \frac{L}{L_0} v|_{L_0},
\quad
\xi = L v|_{L_1}.
\end{eqnarray}
Here nonzero $c_2$ corresponds to the spontaneous breaking of the chiral symmetry in the IR, while the boundary condition at $z=L_0$ corresponds to the explicit breaking of chiral symmetry. The boundary condition at $z=L_1$ is induced by the scalar potential localized on 
$z = L_1$ boundary.
The boundary interaction is
\begin{eqnarray}
{\cal L}_{IR} &=& - a^4 V(\Phi)|_{L_1},
\quad
V(\Phi) = -\frac{1}{2} m_b^2 \text{Tr} |\Phi|^2 + \lambda_b \text{Tr}|\Phi|^4.
\end{eqnarray}
After the symmetry breaking $SU(2)_L \times SU(2)_R \to SU(2)_V$,
vector- and axial-vector gauge bosons 
$V_\mu = (L_\mu + R_\mu)/\sqrt{2}$,
$A_\mu = (L_\mu - R_\mu)/\sqrt{2}$ are expanded into Kaluza-Klein (KK) modes
\begin{eqnarray}
V_\mu = \frac{1}{\sqrt{M_5 L}} \sum_{n=0}^{\infty} f_n^V (z) V^{(n)}(x),
\quad
A_\mu = \frac{1}{\sqrt{M_5 L}} \sum_{n=1}^{\infty} f_n^A (z) A^{(n)} (x),
\end{eqnarray}
where $V^{(n)}$, $A^{(n)}$ correspond to hadronic  vector and axial-vector currents, respectively.

The bulk scalar $\Phi$ is decomposed into $\Phi = (v + \Phi_S) e^{i \Phi_P/v}$
and gauge fixing conditions are
\begin{eqnarray}
\partial_5 (a V_5) =0,
\quad
\Phi_P = -\frac{1}{\sqrt{2} a^3 v} \partial_5 (a A_5),
\end{eqnarray}
and $\Phi_{P,S}(x,z)$ have the following KK expansions: 
\begin{eqnarray}
\Phi_S (x,z) = \frac{1}{\sqrt{M_5 L}} \sum_{n=1}^\infty f_n^S (z) S^{(n)}(x),
\quad
\Phi_P(x,z) = \frac{1}{\sqrt{M_5 L}} \sum_{n=0}^\infty f_n^P (z) P^{(n)}(x), 
\end{eqnarray}
where $S^{(n)}$ and $P^{(n)}$ are scalar and pseudo-scalar hadronic states, respectively.
In particular $P^{(0)}$ corresponds to the pion.

In the QCD the two point correlators for the scalars  and pseudoscalars are
defined as
\begin{eqnarray}
\Pi_{S,P} (p^2) = - \int d^4 x e^{ipx} \langle J_{S,P}(x) J_{S,P}(0)\rangle,
\quad
J_S = \bar{q}q,
\quad
J_P = \bar{q}i\gamma_5 q,
\end{eqnarray}
where the two-point correlator can be obtained from the generating function ${\cal S}$ according to
\begin{eqnarray}
\Pi_S = \frac{\delta^2 {\cal S}}{\delta \hat{s}^2},
\quad
\Pi_P = \frac{\delta^2 {\cal S}}{\delta \hat{p}_s^2},
\end{eqnarray}
where $\hat{s}$ and $\hat{p}_s$ are the scalar and pseudoscalar external sources coupled to QCD:
\begin{eqnarray}
{\cal L} \supset  -\text{Tr} [\bar{q}_L \phi q_R] + h.c.,
\quad
\phi = M_q + \hat{s} + i\hat{p}_s.
\end{eqnarray}

According to the AdS/CFT correspondence, the generating function ${\cal S}$ is obtained by integrating bulk fields restricted to a given Ultraviolet(UV)-boundary value which play the role of the external sources coupled to QCD.
For the 5D scalar field we have
\begin{eqnarray}
\Phi|_{L_0} = \alpha \frac{L_0}{L} \phi,
\end{eqnarray}
($\alpha$ is a constant which will be determined in the matching of correlation in UV as $\alpha = \sqrt{3}$ \cite{DaRold:2005vr}) or explicitly,
\begin{eqnarray}
\Phi_S |_{L_0} = \alpha \frac{L_0}{L} \hat{s},
\quad
\Phi_P|_{L_0} = \alpha \frac{L_0}{L} \hat{p}_s.
\end{eqnarray}
Since the quark masses are given by
$
M_q = \lambda_Q v_S
$,
the singlet scalar fluctuation $s$  
can be related to the scalar source term as
\begin{eqnarray}
\hat{s} \leftrightarrow \lambda_Q s.
\end{eqnarray}
This correspondence can be used to obtain the couplings of $s$ to the meson states.

This AdS/QCD model has five relevant free parameters: $M_q$, $L_1$, $M_5$, $\xi$ and $\lambda$.
$M_q$ is traded with the pion mass $M_\pi$ by the Gell-Mann--Oaks--Renner relation
\begin{eqnarray}
F_\pi^2 M_\pi^2 = - \text{Tr}[M_q] \langle \bar{Q} Q \rangle,
\label{GOR}
\end{eqnarray}
or $M_\pi^2 = - \text{Tr}[M_q] B_0$ where $B_0 \equiv -\langle \bar{Q}Q\rangle/F_\pi^2$.
$M_5$ is related with the beta-function of the QCD and we fix
\begin{eqnarray}
M_5 L &=& \frac{N_{h,c}}{12\pi^2} \equiv \tilde{N}_c, 
\end{eqnarray}
where we consider the case $N_{c,h}=3$.
$L_1$ is related with the mass of the first KK state of $V_\mu$ which corresponds to the 
rho meson mass by $M_{V^{(1)}=\rho} \simeq 2.4/L_1$.
The value of $\xi$ can be fixed by adjusting the mass of the first KK of $A_\mu$
and the first KK vector meson mass
\begin{eqnarray}
\frac{m_{A^{(1)}}}{m_{V^{(1)}}} 
= 
\frac{m_{a_1(1260)}}{m_{\rho(770)}},
\end{eqnarray}
with $m_{a_1(1260)} = 1230 \pm 40~\text{MeV}$ and $m_{\rho(770)}=770~\text{MeV}$, 
which yields \cite{DaRold:2005zs}
\begin{eqnarray}
\xi  \simeq4.
\end{eqnarray}
The pion decay constant $F_\pi$ is written in terms of $L_1$, $\xi$, $\tilde{N}_{c,h}$ as
\begin{eqnarray}
F_\pi^2 &=& \frac{2^{5/3} \pi \tilde{N}_{c}}{3^{1/6} \Gamma(\tfrac{1}{3})^2} \frac{\xi^{2/3}}{L_1^2},
\end{eqnarray}
when $\xi \gg1$. Here $\Gamma(x)$ is the gamma function.
With $L_1 = 320~\text{MeV}$ and $N_{h,c}=3$ one has
\begin{eqnarray}
F_\pi = 87(\xi/4)^{1/3}~\text{MeV},
\end{eqnarray}
which well agree with the experimental value \cite{PDG} $F_{\pi^-} = 130.4\pm0.2~\text{MeV}$
when $\xi \simeq 4$.
Hence for $N_{h,c}=3$ and $\xi=4$ we have
$F_\pi = 0.27 L_1^{-1}$. The hidden rho meson mass $M_{\rho_h} = M_{V_1}$
and hidden axial vector meson mass $M_{a_{1,h}} = M_{A_1}$ is estimated as
\begin{eqnarray}
M_{\rho_h} = 2.4L_1^{-1} = 8.9 F_\pi,
\quad
M_{a_{1,h}} \simeq \frac{1230}{770} M_{\rho_h} \simeq 14.1 F_\pi.
\end{eqnarray}

\begin{figure}[htbp]
\includegraphics[width=7cm]{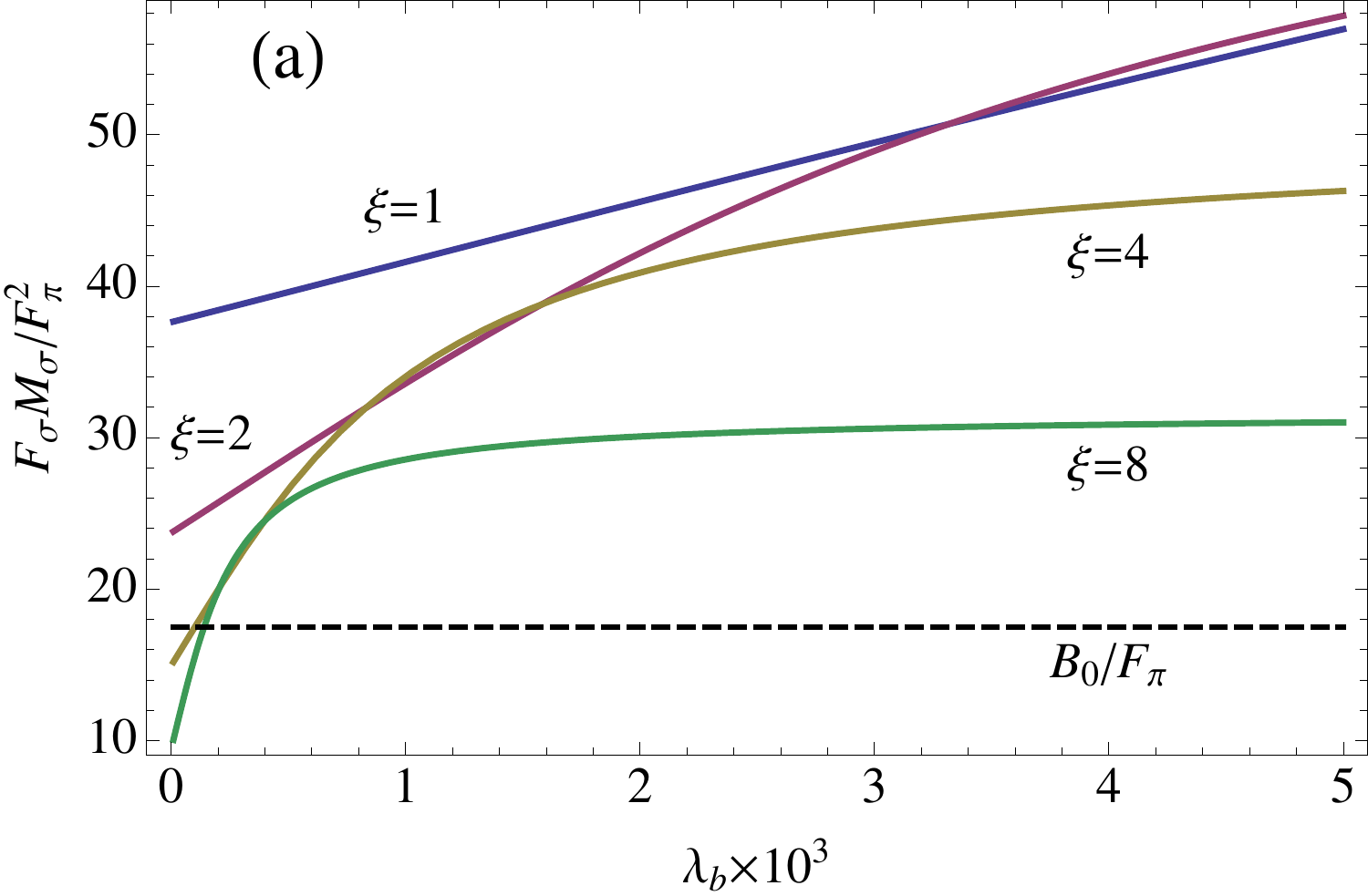}
\includegraphics[width=7cm]{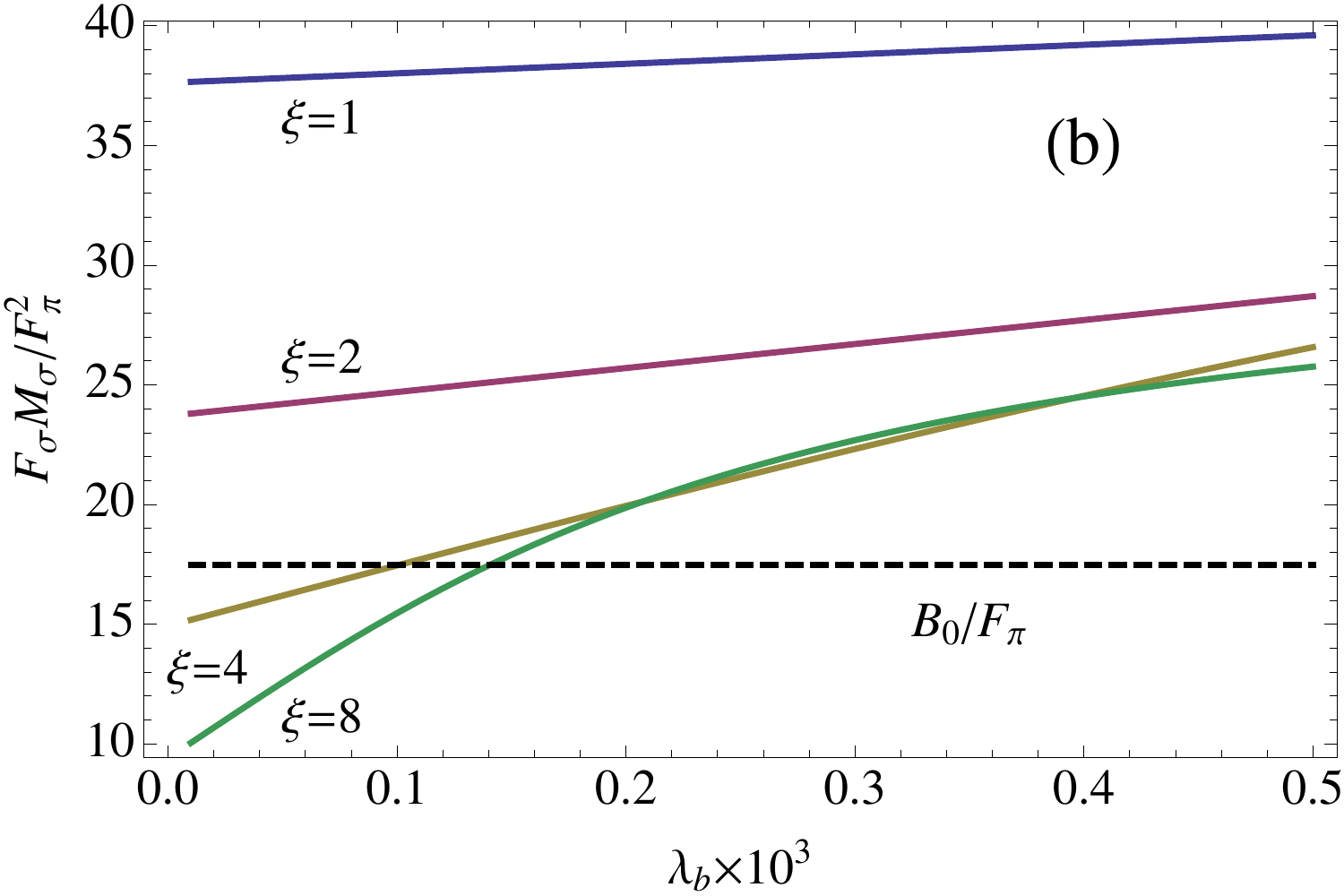}
\caption{
$F_\sigma M_{\sigma}$ in the unit of $F_\pi^2$ 
for $\xi=1,2,4,8$, $N_{h,c}=3$.
$B_0/F_\pi$ is indicated as black dashed line.
(a) $0 \le \lambda_b \le 5\times 10^{-3}$
 (b) $0 \le \lambda_b \le 5 \times 10^{-4}$.
}\label{fig-FM}
\end{figure}

Now we fix the value of $\lambda_b$.
In the original paper the author estimated $\lambda_b = 10^{-2} - 10^{-3}$ and identified
the lightest scalar meson as $a_0(980)$. 
In the present study, we regard  sigma meson as lightest scalar resonance state, $S^{(1)}=\sigma$.
In the AdS/QCD, since the wave functions of $S^{(n)}$, $P^{(n)}$ vanish at $z=L_0$
\begin{eqnarray}
f_n^{S,P} |_{z=L_0} = 0,
\end{eqnarray}
there are no direct interactions between meson states with source term.  
In the AdS/QCD, source-pion-pion interactions are given by
\begin{eqnarray}
{\cal L}_{\pi^2\hat{s}} 
&=& - \tilde{B}_0 \text{Tr} [\pi^2 \hat{s}] +
  \text{Tr} [(\partial_\mu\pi)^2 \hat{s}] 
\sum_n  \frac{{G_{n\pi\pi}} F_{S_n} M_{S_n}}{p^2 + M_{S_n}^2},
\label{source-int}
\end{eqnarray}
($B_0 F_\pi^2 \equiv - \langle \bar{Q}Q\rangle$)
where $M_{S_n}$, $F_{S_n}$ and $G_{n\pi\pi}$ is the mass, decay constant and 
the $S_n (\partial_\mu\pi)^2$ coupling.
These terms arise due to the $S_n - \hat{s}$ mixing.
In particular, the first term of r.h.s. of eq. \eqref{source-int}
is induced by $\sigma-\pi-\pi$ coupling through the $\sigma$-source mixing.
We assume that $F_{S_1} = F_\sigma$, $M_{S_1} = M_{\sigma\sigma}$ and that
the mixing is given by
\begin{eqnarray}
F_\sigma M_{\sigma\sigma} = m_{S\sigma}^2.
\end{eqnarray}
Together with the Gell-Mann--Oaks--Renner  relation eq. \eqref{GOR}
we obtain
\begin{eqnarray}
F_{\sigma} M_{\sigma\sigma} = B_0 F_\pi.
\label{adsqcd-rel}
\end{eqnarray}
In the AdS/QCD, 
\begin{eqnarray}
\langle \bar{Q} Q \rangle = - 2\sqrt{3} \tilde{N}_{h,c} \frac{\xi}{L_1^3}
\end{eqnarray}
and $\langle \bar{Q} Q \rangle \simeq - 18 F_\pi^3$ is obtained for $N_{h,c}=3$, $\xi=4$.
$M_{S_n}$ and $F_{S_n}$ is obtained by formulas summarized in the Appendix.
We find numerically that eq. \eqref{adsqcd-rel} is satisfied when
\begin{eqnarray}
\lambda_b \simeq 1.0 \times 10^{-4},
\end{eqnarray}
(See fig. \ref{fig-FM}),
and hence we obtain a relation (See fig. 2)
\begin{eqnarray}
M_{\sigma\sigma} \simeq 5.0 F_{\pi}
\quad
(\xi_\sigma \simeq 5),
\quad 
F_\sigma \simeq 3.5 F_{\pi}.
\end{eqnarray}

\begin{figure}[htbp]
\includegraphics[width=7cm]{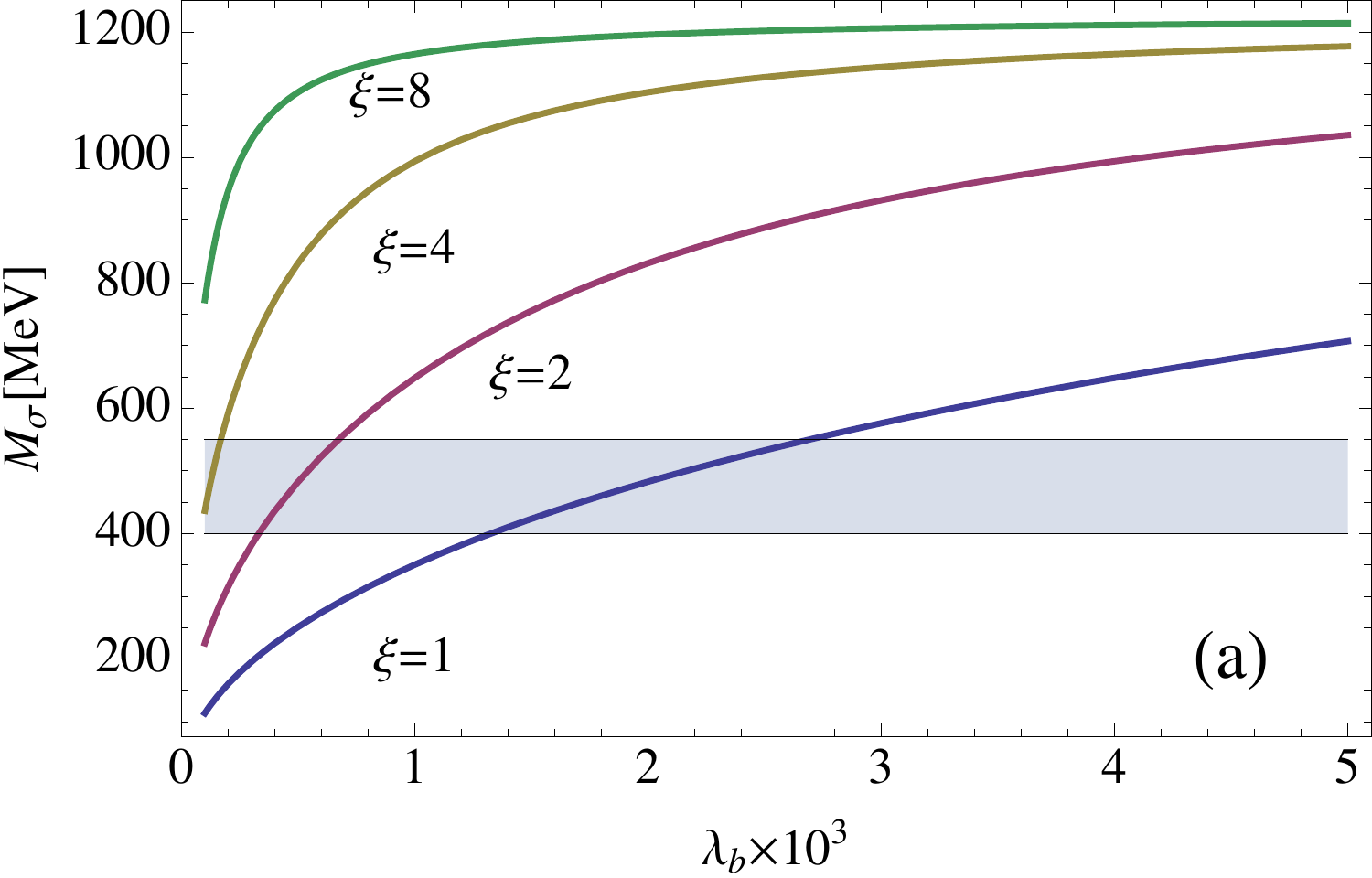}
\includegraphics[width=7cm]{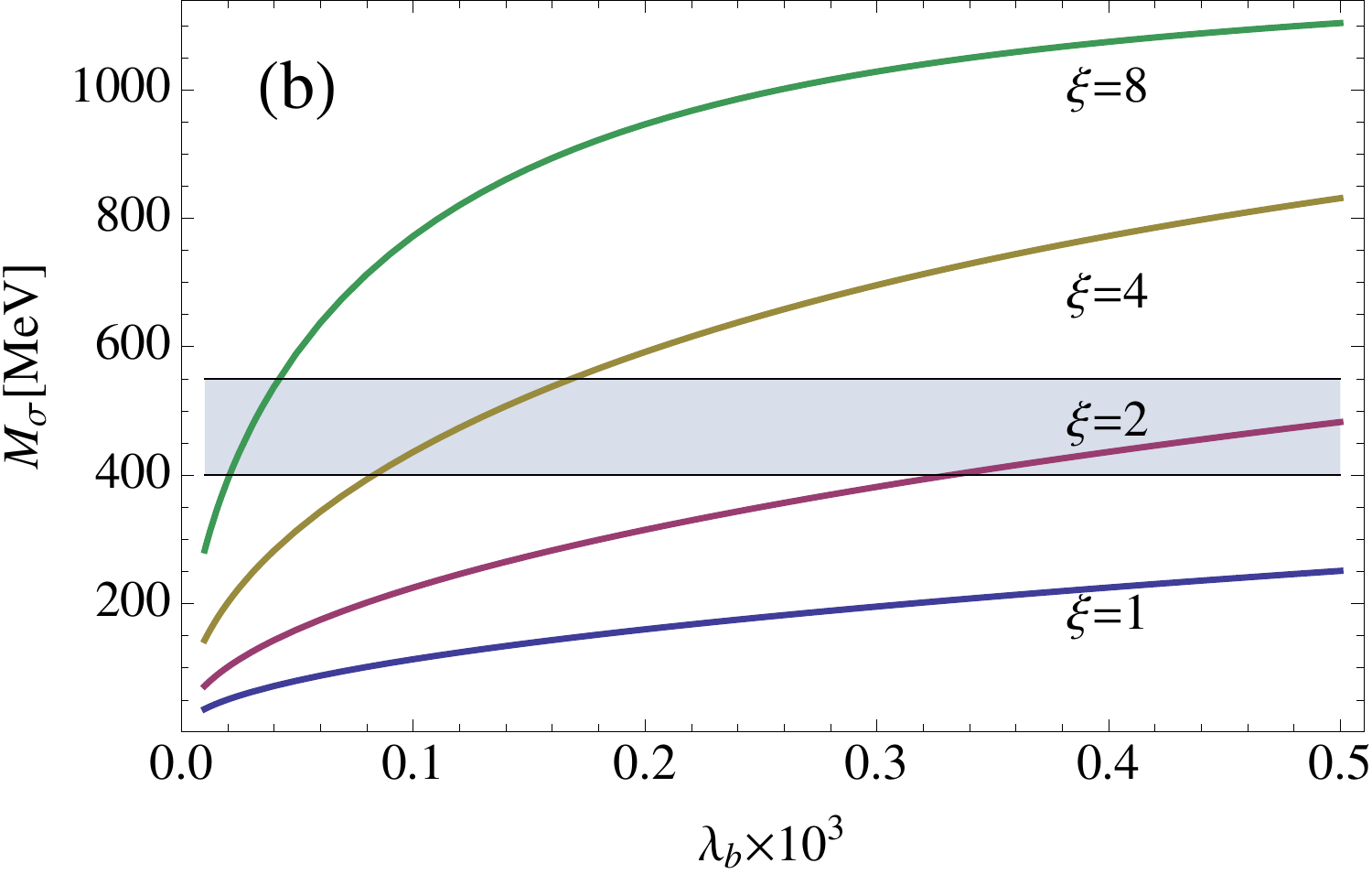}
\caption{
The lightest scalar mass as the sigma meson mass $M_\sigma$ v.s. scalar boundary coupling $\lambda_b$ with $\xi=1,2,4,8$, $N_{h,c}=3$,
$L_1^{-1} = 320~\text{MeV}$.
The shaded band indicates the mass range of $f_0(500)$ \cite{PDG}.
(a) $0 \le \lambda_b \le 5\times 10^{-3}$
 (b) $0 \le \lambda_b \le 5 \times 10^{-4}$.
}\label{fig-Msigma}
\end{figure}

When we take a scale normalized by $L_1^{-1} = 320~\text{MeV}$
so that we have $M_{V_1} = 2.4/L_1=  m_{\rho(770)} = 770~\text{MeV}$,
we obtain $F_{\pi} = 87~\text{MeV}$, $M_\sigma\sim 450~\text{MeV}$ (fig. \ref{fig-Msigma}).
This result well agree with the experimental bound 
$400~\text{MeV} \le m_{f_0(500)} \le 550~\text{MeV}$ \cite{PDG},
if we identify $f_0(500)$ as the sigma meson.

\section{Hidden Pion Phenomenology}\label{sect-dm}
As a result of the previous sections, we now have three free parameters in this  model:
$M_{\pi}, F_\pi$ and $v_S=v_H \tan\beta$. 
For numerical analysis, we scan the three-dimensional parameter space 
$(M_{\pi},F_{\pi},\tan\beta)$. Note that the SM-like Higgs boson with mass $125$ GeV 
is termed as $H$, and extra scalar particles as $H_1$ and $H_2$ with $M_{H1} < M_{H2}$.  
We have considered several theoretical and experimental constraints.

Let us first consider the theoretical constraints. From the stability of the potential, 
the dimensionless couplings should satisfy the 
relation, 
\begin{equation}
 \lambda_{HS} \leq \lambda_{H}\cdot\lambda_{S}~~{\rm with} ~~\lambda_H,\lambda_S >0,
\end{equation}
which are translated into the following relation with the minimization conditions 
\begin{equation}
\tan\beta \leq \sqrt{2}~\frac{M_\pi F_\pi}{v_H^2}.
\label{tanbrange}
\end{equation}
Since the strongly interacting hidden sector are now treated as linear sigma model, 
the symmetry breaking condition in the $\sigma$-sector would constrain 
the value of  $\mu^2_\sigma$ to be positive. With the help of 
eq. \eqref{positivemu2}, the condition reads
\begin{equation}
M_\pi \leq \frac{\xi_\sigma F_\pi}{\sqrt{3}}.
\label{mu2range}
\end{equation}
We also adopt the perturbativity bound on $\lambda_S$ with our definition of the 
Lagrangian: 
\begin{equation}
\lambda_S \leq \frac{4\pi}{3}.
\end{equation}

Experimental constraints considered in the analysis are listed in the following:
\begin{itemize}
\item Signal strength for the SM Higgs boson 
\cite{Khachatryan:2014jba,Aad:2015gba},
\begin{equation} 
\hat{\mu}=1.00\pm0.13 ~\text{\cite{Khachatryan:2014jba}}. 
\end{equation}

\item Bounds for extra scalar particles from the LEP \cite{Barate:2003sz} 
and the LHC \cite{CMS:2012bea,TheATLAScollaboration:2013zha,ATLAS:2014aga}.

\item Relic density from Planck satellite \cite{Ade:2015xua}.
\begin{equation}
\Omega_{DM}h^2=0.1198\pm0.0015.
\end{equation}
\item Neutrino signals through the DM capture by the Sun, mostly  from 
Super-Kamiokande for upward muon flux \cite{Tanaka:2011uf}.
\item Fermi-LAT 6-year results for DM annihilation \cite{Ackermann:2015zua}.
\item Higgs invisible width from the LHC \cite{Aad:2014iia,Chatrchyan:2014tja},
\begin{equation}
{\rm Br}_{inv} \lesssim  0.51,~~{\rm with }~~ 95\%~ {\rm C.L.}
\end{equation}
\item Direct detection bound, mostly from LUX \cite{Akerib:2013tjd}, 
SuperCDMS \cite{Agnese:2014aze}, and CRESST-II \cite{Angloher:2014myn}.
\end{itemize}

 We apply $2 \sigma$ bounds with these experimental constraints except for the
relic density, for which we use the measured value as an upper bound.  
 This is because 
there could be additional contributions from hidden baryons to DM thermal relic density,
which we do not include in this paper. 
We vary the $F_\pi$ up to 2 TeV and the ranges for $M_\pi$ and
$\tan\beta$ are fixed with eq. \eqref{tanbrange}  and eq. \eqref{mu2range}.
We use {\tt micrOMEGAs} \cite{Belanger:2013oya} for evaluating DM-related
observables.

\begin{figure}[t]
\includegraphics[width=8cm]{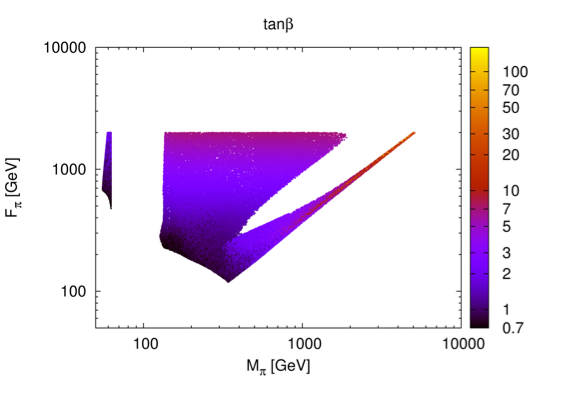}
\caption{Solution points are shown in $(M_\pi,F_\pi)$ plane with color contours for 
$\tan\beta$.}
\label{fig3}
\end{figure}

The result of scanning is depicted in fig. \ref{fig3}.   Here we see that there is
 definite lower bound for $F_\pi$, around 100 GeV.  Also $\tan\beta$ is  bounded 
from below, $\tan\beta \gtrsim 0.7 $, mainly because of the perturbativity of 
$\lambda_S$ since it  can be written in the form
\begin{equation}
\lambda_S=\frac{\lambda_{HS}}{\tan^2\beta}
 +\frac{2 M_\pi^2 F_\pi^2}{v_H^4 \tan^4\beta}.
\end{equation}
So if $\tan\beta$ is too small, $\lambda_S$ will have very large value, above 
the perturbativity bound.  The island on the leftmost side are the solution 
points where $2 M_\pi \sim M_{H}$, {\it i.e.} the SM-like Higgs resonances. 
Other points include light scalar resonanaces with $2 M_\pi \sim M_{H1}$, 
heavy scalar resonances with $2 M_\pi \sim M_{H2}$ and non-resonance solutions
 with $M_{H1} \ll 2 M_\pi \ll M_{H2}$. The non-resonance solutions favor 
relatively small $\tan\beta$. For example, if the hidden pion mass is away from 
 both resonance regions more than $30\%$, {\it i.e.}  
$1.3 \cdot M_{H1} \leq 2 M_\pi \leq 0.7 \cdot M_{H2}$, 
then $\tan\beta$ is constrained to be smaller than about $3$. 
No points survive if the hidden pion mass is far away from the resonance regions, 
more than about $50\%$.
This is because when $\tan\beta$ 
is small, the off-diagonal term of the mass matrix presented in eq. \eqref{mmatrix},
$M^2_{hs}=-M_{hh}^2/\tan\beta$, is 
enhanced compared with large $\tan\beta$ case so  that the mixing between 
the SM-like Higgs boson and singlet scalar fields are enhanced.

\begin{figure}[t]
\includegraphics[width=8cm]{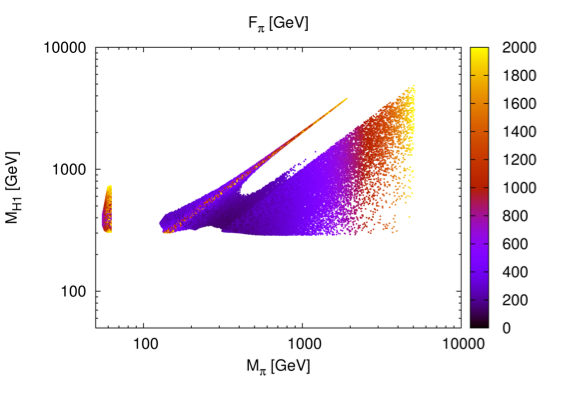}
\includegraphics[width=8cm]{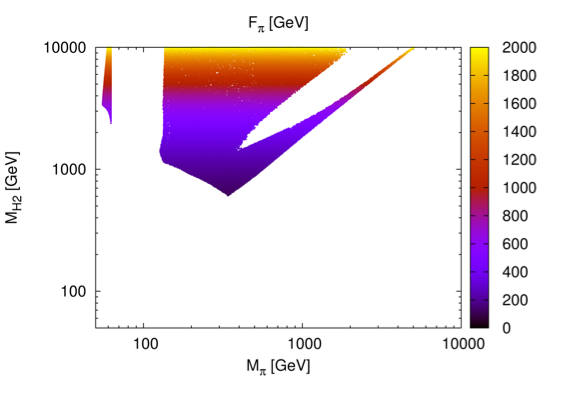}
\caption{Solution points are shown in $(M_\pi,M_{H1})$ plane (left) 
and $(M_\pi,M_{H2})$ plane (right), with color contours for the hidden pion decay constant $F_\pi$.}
\label{fig4}
\end{figure}

These features for the solution points can be more easily checked with 
another forms of plots.
 In fig. \ref{fig4}, the solution points are shown in $( M_\pi, M_{H1,H2} )$ 
plane.
The thin branch in the left plot is the collection of solution points where 
$M_\pi \sim M_{H1}/2$, light scalar resonances. Note that there is no 
solution points when $M_{H1} < M_H$, {\it i.e.} all extra scalar particles 
are heavier than the SM-like Higgs boson. The right plot, 
where the points are shown in $(M_\pi,M_{H2})$ plane, also includes 
the thin branch that is corresponding to the solution points with heavy scalar 
resonances. The shape of the plot is almost same as 
fig. \ref{fig3}, since a relation  
$M_{H2} \sim \xi_\sigma F_\pi \sim 5 F_\pi$ generally holds in this model. 
As a result, $M_{H2}$ has a definite lower bound ${\color{black} M_{H2} \gtrsim 590~{\rm GeV}}$ 
as ${\color{black} F_\pi \gtrsim 115~{\rm GeV}}$ does. 
Both plots also include the non-resonance cases, 
of which DM mass are far from both resonances and other parameters are tuned 
to satisfy the all constraints.

\begin{figure}[t]
\includegraphics[width=8cm]{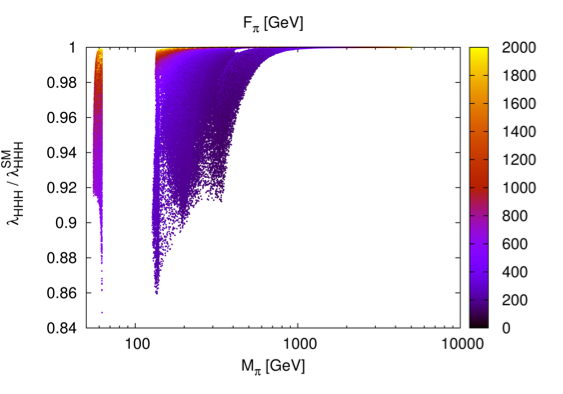}
\caption{Triple Higgs couplings normalized with the SM values versus the hidden pion masses. 
Color contours represent the hidden pion decay constant $F_pi$.}
\label{fig5}
\end{figure}

As a distinctive observable, we show the deviation of the triple Higgs 
coupling from 
the SM prediction in fig. \ref{fig5}. 
The triple Higgs coupling in the model can reach $\sim 85 \%$ of the 
SM prediction for relatively small $M_\pi$. 
If we take larger values for $M_\pi$ and $F_\pi$, it 
approaches to the SM one and cannot be a distinctive observable. 
Especially when $M_\pi$ is larger than 1 TeV, the triple Higgs coupling is 
very close to the SM prediction and the deviation cannot be detected.

\begin{figure}[tb]
\includegraphics[width=8cm]{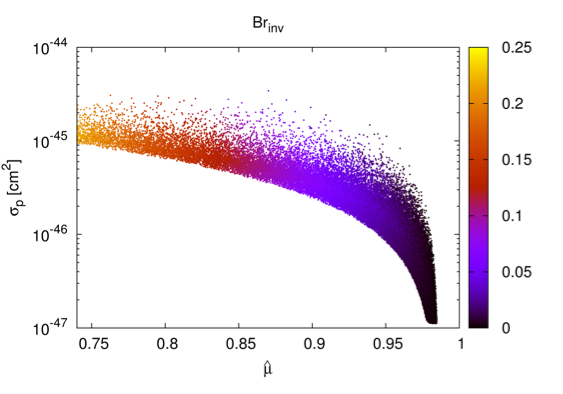}
\includegraphics[width=8cm]{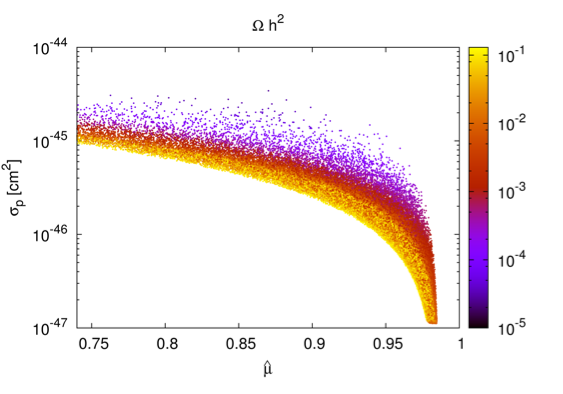}
\caption{Correlation between the DM-nucleon cross sections and the Higgs signal strengths in 
the SM-like Higgs resonance cases. 
Color contours represent the SM-like Higgs invisible decay withs (left)
 and thermal relic density of the DM (right).}
\label{fig6}
\end{figure}

Let us consider another observables. To be more specific, we separate the 
cases as the SM-like Higgs resonance and the other cases. 
fig. \ref{fig6} shows the correlations 
between the Higgs signal strength $\hat{\mu}$ and DM-nucleon cross section in 
Higgs resonance case.  Deviation of $\hat{\mu}$ from 1 is generated by the 
mixing angle $|V_{h0}|^2$ in eq. \eqref{mixingmat}. In this case, 
the SM-like Higgs boson can decay to 
a pair of DM's and these decay modes contribute to the Higgs invisible decay 
width. As one can see in the left plot, 
the invisible decay width of the SM-like Higgs boson increases when 
 the signal strength decreases and vice versa, just like DM-nucleon 
cross section. 
We can understand this by the fact that the deviation of 
the signal strength is determined entirely by non-zero mixing angles among the 
SM-like Higgs and other extra scalar particles.  The more they are mixed,  the more 
wide the channel between the visible and hidden sector is open. 
We show in the right plot the same correlation with relic density contours. 
The variation of the relic density is caused by the small variation of 
$M_\pi$ and $F_\pi$ as long as $M_\pi$ is close to $\sim M_{H}/2$. In both 
plots we apply $2 \sigma $ bound for the signal strength $\hat{\mu}$ 
such as $\hat{\mu} \ge 0.74$.

\begin{figure}[t]
\includegraphics[width=8cm]{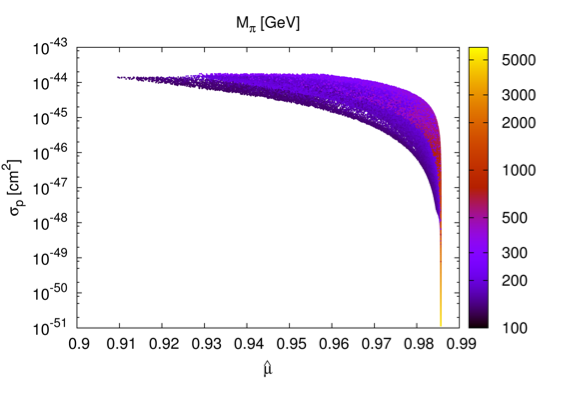}
\caption{Correlation between the DM-nucleon cross sections and the Higgs signal strengths in 
resonance and non-resonance cases ,except the SM-like Higgs resonances. 
Color contours represent the hidden pion masses.}
\label{fig7}
\end{figure}

The same correlation is shown in fig. \ref{fig7} for the other cases than the 
SM-like Higgs resonance. As mentioned before, the cases include light and 
heavy scalar resonances and non-resonance solutions. In this plot, the color 
contour represents the mass of the DM. Unlike the Higgs resonance case, 
relatively large values of the signal strength are favored, with generically 
larger values of DM-nucleon cross section. Note that there is an upper 
limit for $\hat{\mu}$. This bound is originated from the $h-\sigma$ mixing, 
that should be different from zero for avoiding the overclosure of the 
universe by the DM.

\begin{figure}[t]
\includegraphics[width=8cm]{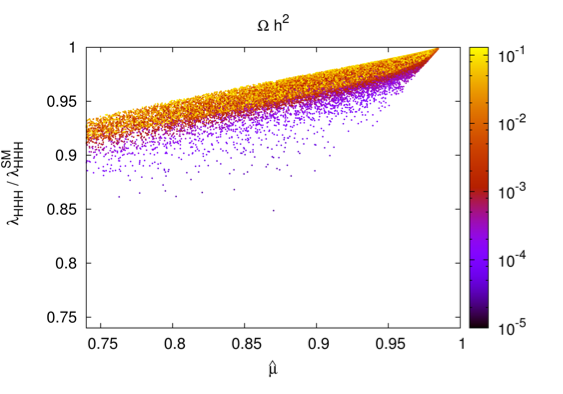}
\includegraphics[width=8cm]{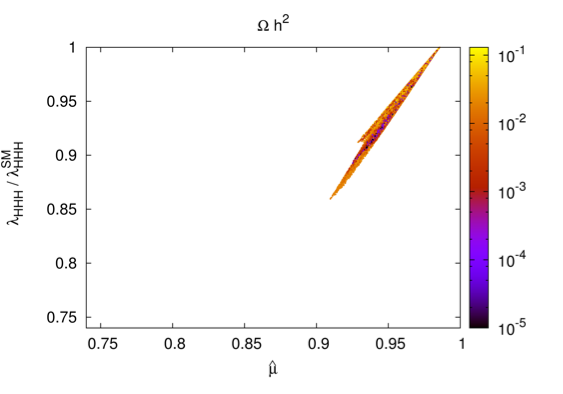}
\caption{Correlations between the Higgs signal strength and triple Higgs couplings normalized with 
the SM values for the SM-like Higgs resonance (left) and the other cases (right). Color contours 
represent thermal relic densities.}
\label{fig8}
\end{figure}

fig. \ref{fig8} is showing the correlation between the signal strength and 
triple Higgs coupling normalized with the SM prediction. The left plot is 
for the SM-like Higgs resonance solutions and the right one for other cases. 
Both cases are predicting sharp linear correlations, but with different slopes.
 Two solution points are disjoint with each other, so the measurements of them 
can be used for the identification of the scenarios, though their values are 
small.

Finally, some benchmark points are collected in Table \ref{table1}.
We classify the points as (A) SM Higgs resonance solutions, (B) light scalar
resonances, (C) heavy scalar resonances, and (D) non-resonance solutions.
The solution points with large relic densities
(close to upper bound, $\Omega_{DM}^{exp} h^2 \sim 0.1198$) are labelled
as (I) and (II,III) correspond to the cases of small relic densities. 
Note that the closer we take the DM mass to the exact resonance, $M_{H}$
(or $M_{H1}, M_{H2}$), the smaller the relic density becomes.

\begin{center}
\begin{table}
\begin{tabular}{|c|c|c|c|c|c|c|c|c|c|c|}
\hline
\hline
type & $F_{\pi}/{\tiny \rm GeV}$ & $M_{\pi}/{\tiny \rm GeV}$ & $\tan\beta$ & $M_{H1}/{\tiny \rm GeV}$ & $M_{H2}/{\tiny \rm GeV}$ & $\Omega h^2$ & 
$\sigma_p/{\tiny \rm cm^2}$ & $\hat\mu$ & ${\rm BR_{Inv}}$ & ${\lambda_{HHH}}/{\lambda_{HHH}^{SM}}$ \\ 
\hline 
\hline
A-I & 692.19 & 55.27  & 0.7881 & 346.281 & 3460.96 & 0.1220 & $9.86\cdot 10^{-46}$  & 0.754 &  0.206  & 0.932  \\
\hline
A-II & 1524.33 & 62.49  & 1.648 & 415.13 & 7621.67 & 0.0027 & $3.29\cdot 10^{-47}$  & 0.981 &  $3.8\cdot 10^{-4}$  & 0.994  \\
\hline
\hline
B-I & 1244.17 & 1000.08 & 4.288 & 2033.54 & 6224.10 & 0.1088 & $2.88\cdot10^{-49}$ & 0.986 & 0 & 1.000 \\
\hline
B-II & 1607.03 & 999.63 & 5.638 & 2000.65 & 8036.53 & 0.0101 & $1.05\cdot10^{-49}$ & 0.986 & 0 & 1.000 \\
\hline
B-III & 775.72 & 199.22 & 2.761 & 396.81 & 3878.60 & 0.0061 & $4.92\cdot10^{-47}$ & 0.984 & 0 & 0.998 \\
\hline
\hline
C-I & 227.49 & 549.85 & 1.479 & 570.38 & 1153.54 & 0.1193 & $6.08\cdot10^{-46}$ & 0.985 & 0 & 0.988 \\
\hline
C-II & 387.01 & 999.03 & 6.867 & 378.17 & 1938.82 & 0.0681 & $2.07\cdot10^{-47}$ & 0.985 & 0 & 1.000 \\
\hline
\hline
D-I & 185.66 & 319.49 & 1.337 & 323.97 & 930.67 & 0.1192 & $4.94\cdot10^{-45}$ & 0.964 & 0 & 0.965 \\
\hline
D-II & 208.73 & 231.86 & 0.906 & 405.30 & 1044.98 & 0.0101 & $6.50\cdot10^{-45}$ & 0.967 & 0 & 0.966 \\
\hline
\hline
\end{tabular}
\caption{Benchmark points for (A) SM-like Higgs resonance, 
 (B) light scalar resonance, (C) heavy scalar resonance and (D) non-resonance cases.}
\label{table1}
\end{table}
\end{center}

\section{Summary and Conclusion}\label{sect-summary}

In this paper, we have analyzed the scale-invariant extension of the SM with 
vector-like confining gauge theory in the hidden sector by using the AdS/QCD 
proposed in Refs. \cite{DaRold:2005zs,DaRold:2005vr}. The model 
contains the singlet scalar field that connects the confining hidden sector 
and the scale-invariant SM sector. Hidden sector fermions develop nonzero chiral 
condensates and generate the linear term in the potential of the singlet scalar field $S$. 
As a result,  the singlet scalar field  $S$  develops a nonzero VEV and it provides 
the tachyonic  mass term for the SM Higgs field. Therefore the origin of the EWSB in the 
SM sector lies in  the new strong dynamics in the hidden sector.  

We have used the AdS/QCD approach to describe non-perturbative 
dynamics of the hidden QCD sector. 
By the AdS/QCD, strongly interacting $SU(3)$ gauge theory in the hidden sector 
with two-flavors can be described by $SU(2)_L \times SU(2)_R$ gauge theory 
on $AdS_5$. The spectrum of the mesonic states then can be calculated up to 
overall scale by considering the two-point correlators. 
We {\color{black} first fixed} the values of the AdS/QCD parameters that 
reproduce the known spectra of the mesons by identifying first KK mode of 
the vector state as rho meson. We applied the results to the hidden 
QCD. In this case, hidden rho meson mass is be treated as  
overall scale of the hidden QCD.  By this, we successfully found out the relation between 
hidden sigma meson mass and hidden pion decay constant etc. 
As a result, we reduced the number of free parameters of the model to 
three, {\it i.e.} $F_\pi, M_\pi$ and $\tan\beta$.   

The hidden pions can be the DM candidates since the hidden sector flavor symmetry 
becomes an accidental symmetry of hidden sector strong interaction.
We have analyzed these ``hidden pion''
properties as the DM. Many results of the 
DM search experiments were considered. In addition to the SM-like Higgs boson, 
we have two extra neutral scalar 
fields in the model. Those extended scalar sectors are constrained by the 
LHC data, for example Higgs signal strengths and non-observation of another scalar 
particles, etc. By scanning the three-dimensional parameter space 
$(M_\pi, F_\pi, \tan\beta)$, we found that the non-resonance solutions are 
also possible in addition to the resonance solutions.  We also considered 
various correlations among the experimental observables. For example, there 
is the correlation between the Higgs signal strength $\hat{\mu}$ and DM-nucleon cross 
section, and also between $\hat{\mu}$ and the triple SM-like Higgs coupling. 
Especially for the latter,  we found that their values and correlations behave differently 
depending on whether hidden pions have the SM-like Higgs resonance or not.
Though the Higgs signal strength $\hat{\mu}$ has been measured quite  
precisely and seems to be consistent with the SM prediction, there is 
still room for the physics beyond the SM as discussed in this paper. 
If the Higgs signal strength $\hat{\mu}$ is measured  more precisely, according to 
the sharp correlations we found, we can give  peculiar predictions on the DM properties 
and others such as triple Higgs coupling etc. This could be seen in the benchmark points 
we presented at the  end of the analysis. 

Let us comment on the future prospects. Our model contains two extra neutral 
scalar bosons that mix with the SM-like Higgs boson. Mass spectra 
of those two scalar bosons are constrained by the up-to-date experimental 
results {\color{black} on the Higgs signal strengths} in such a way that both of them are 
heavier than the 125 GeV SM-like Higgs boson. 
Besides the resonance solutions by the SM-like Higgs, extra 
light and heavy scalar particles, the non-resonance solutions are also 
possible for moderate values of hidden pion mass and decay constant. 
In that case, the mass of the light extra scalar particle
 will be around a few hundred GeV, which can be accessible 
at the LHC Run-II. The model also predicts the values of other 
observables such as relic density, DM-nucleon cross section and triple Higgs coupling 
and so forth. Especially, the Higgs signal strength $\hat{\mu}$ will be sharply determined. 
The more detailed study on the collider phenomenologies, for example, 
the pair production of the SM-like Higgs boson, could be possible. 
In addition, more complete studies with the hidden baryons, another DM 
candidates, can be pursued with the AdS/QCD.

\begin{acknowledgments}
This work is supported in part by 
National Research Foundation of Korea (NRF) Research 
Grant NRF-2015R1A2A1A05001869 (HH, DWJ, PK),  
NRF-2015R1D1A1A01059141, NRF-2015R1A2A1A15054533 (DWJ)
and by the NRF grant funded by the Korea government (MSIP) 
(No. 2009-0083526) through Korea Neutrino Research 
Center at Seoul National University (PK). 
\end{acknowledgments}

\appendix

\section{AdS/QCD formulas}\label{sectA-formula}

Scalar meson mass $M_{S_n}$ and decay constant $F_{S_n}$ are obtained 
\cite{DaRold:2005vr} by comparing the 
scalar correlator
\begin{eqnarray}
\Pi_S(p_E^2) &=& \alpha^2 M_5 L \left[
\frac{1}{L_0^2} + \frac{ip_E}{L_0}
\frac{J_0(ip_E L_0) + b(p_E Y_0 Y_0(ip_E L_0))}{J_1(ip_E L_0) + b(p_E) Y_1(ip_E L_0)}
\right] 
\\
&\stackrel{L_0\to0}{\simeq}&
\alpha^2 M_5 L \left[ \frac{1}{L_0^2} + \frac{1}{2}p_E^2 \ln (p_E^2 L_0^2)
+ \frac{\pi p_E^2}{2 b(p_E)}\right],
\\
b(p_E) &=&
 - \frac{
 ip_E L_1 J_2(ip_E L_1) - \frac{8\lambda_b \xi^2}{M_5L} J_1(ip_E L_1)
 }{
 ip_E L_1 Y_2(ip_E L_1) - \frac{8\lambda_b \xi^2}{M_5L} Y_1(ip_E L_1)
}
\end{eqnarray}
(where $p_E$ is the Euclidean momentum) with the correlator in Large-$N$ QCD
\begin{eqnarray}
\Pi_S(p_E^2) &=& \sum_n \frac{F_{S_n}^2 M_{S_n}^2}{p_E^2 + M_{S_n}^2}.
\end{eqnarray}
The masses of scalar resonances are determined by finding the poles, $b(p_E)=0$
or
\begin{eqnarray}
M_{S_n} L_1 J_2(M_{S_n} L_1) = \frac{8\lambda_b\xi^2}{M_5L}J_1(M_{S_n} L_1),
\end{eqnarray}
and corresponding residues gives the scalar decay constants
\begin{eqnarray}
F_{S_n}^2 &=& \frac{3\tilde{N}_c \pi M_{S_n}^2 (
\frac{8\lambda_b\xi^2}{M_5L} Y_1(M_{S_n}L_1) - M_{S_n}L_1 Y_2(M_{S_n}L_1)
)}{
M_{S_n} L_1 (1-\frac{8\lambda_b\xi^2}{M_5L}) J_0(M_{S_n}L_1) + (\frac{8\lambda_b\xi^2}{M_5L} + M_{S_n}^2 L_1^2 - 2) J_1(M_{S_n}L_1)
}.
\end{eqnarray}


  
\end{document}